\documentclass[journal,10pt]{IEEEtran}
\usepackage{geometry}
\usepackage{graphicx}
\usepackage[fleqn]{amsmath}
\usepackage{cases}
\usepackage{subfigure}
\usepackage{stfloats}
\usepackage{epstopdf}
\usepackage{cite}
\usepackage{algorithm}
\usepackage{algorithmic}
\usepackage{tabu}

\usepackage{amssymb}
\usepackage{color}
\usepackage{xcolor}

\begin{document}
\title{Kalman-Like Filter under Binary Sensors}

\author{Zhongyao~Hu, Bo~Chen, Yuchen Zhang, Li~Yu
\thanks{Z. Hu, B. Chen, W. Zhang and L. Yu are with Department of Automation, Zhejiang University of Technology, Hangzhou 310023, China. (email:
bchen@aliyun.com).}
}

\maketitle

\begin{abstract}
This paper is concerned with the linear/nonlinear Kalman-like filtering problem under binary sensors. Since innovation represents new information in the sensor measurement and serves to correct the prediction for the Kalman-like filter (KLF), a novel uncertain measurement model is proposed such that the innovation generated from binary sensor can be captured. When considering linear dynamic systems, a conservative estimation error covariance with adjustable parameters is constructed by matrix inequality, and then an optimal filter gain is derived by minimizing its trace. Meanwhile, the optimal selection criterion of an adjustable parameter is developed by minimizing the upper bound of the conservative estimation error covariance. When considering nonlinear dynamic systems, a conservative estimation error covariance with adjustable parameters is also constructed via unscented transform and matrix inequalities. Then, following the idea of designing KLF in linear dynamic systems, the nonlinear filter gain and the optimal adjustable parameter are designed. Finally, $O_2$ content estimation and nonlinear numerical system are employed to show the effectiveness and advantages of the proposed methods.
\end{abstract}


\vspace{-3pt}
\section{Introduction}
Kalman filter [1] is a minimum mean square error estimator for linear Gaussian systems, which have the advantages of high accuracy, low computational effort and ease of implementation. In fact, a large number of practical systems are nonlinear, which limits the application of the Kalman filter. In this case, a series of Kalman-like filters (KLFs) such as extended Kalman filter [2], unscented Kalman filter [3] and cubature Kalman filter [4] have been proposed, and these nonlinear filters have the similar structure of Kalman filter. Notice that the above methods are developed based on traditional continuous-value sensors which transmit sensor data completely. However, sensor energy and bandwidth constraints in the communication environment are frequent problems, and thus under these situations sensor data cannot be transmitted completely. To overcome the above-mentioned problems, different KLFs have been proposed based on quantization method [5-8] and dimensionality reduction method [9-11], and these methods can reduce the size of the sensor data being transmitted. At the same time, binary sensors are a special type of sensors that output one bit of data by comparing their sensed variables and thresholds. Obviously, binary sensors can minimize the size of the data being transmitted, and thus the problems of energy and bandwidth limitations are naturally avoided. Particularly, binary sensors are cost-effective, which makes them very flexible in the application, i.e., different performance requirements can be met by arranging different numbers of binary sensors.

Recently, binary sensors have received more and more attention. Medicine [12], internet of things [13], source location [14-15] and many other fields have begun to use binary sensors instead of traditional continuous-value sensors. Due to only 1 bit of information available, state estimation using binary sensors is a very tricky business. To deal with the high nonlinearity of the binary output (similar to the step signal), most of the existing filtering methods under binary sensors were designed based on particle filter (PF) [16], such as the target tracking algorithms in [17]-[18]. However, the PF is computationally intensive and suffers from the curse of dimensionality, which shall be avoided in the proposed methods in this paper. On the other hand, a class of threshold-based methods for processing binary outputs has also received a great deal of attention. This method can extract useful information from binary sensors by analysing their intrinsic measurement form, and then avoid dealing with the high nonlinearity of the binary outputs directly. In [19] and [20], the thresholds of binary sensors were modelled as convex combinations of two sensed variables with uncertainties, and then, moving horizon estimation (MHE) and distributed fusion KLF (DFKLF) for binary sensors were proposed in [19] and [20] based on the threshold model. Subsequently, [21] extended DFKLF to the case where the statistical properties of the system noise were unknown, but it is computationally intensive as the optimization algorithm
is needed to solve the linear matrix inequality. It should be pointed that the above threshold-based methods all have two common shortcomings: 1) The uncertainties in the model must be ignored when designing filters, which may reduce their estimation performance. 2) These methods are only applicable to the linear dynamic systems, but many practical systems are nonlinear.

Motivated by the above analysis, we shall study the Kalman-like filtering problem for linear/nonlinear dynamic systems under binary sensors.
The main contributions of this paper are summarized as follows:
\begin{itemize}
\item[(i)]A novel innovation-based uncertainty model for binary sensor is developed, which can extract more useful information for KLFs than the switch-based uncertainty model in [19-21]. Meanwhile, when the novel model is employed to construct KLFs, the uncertainties induced by binary sensors can be offset rather than ignored directly as in [19-21].
\item[(ii)]
    Based on the proposed new model, KLFs are designed for both linear and nonlinear dynamic systems under binary sensors, and the filter gains are obtained by minimizing the traces of estimation error covariances. Meanwhile, by minimizing the upper bound of the estimation error covariances, an optimal selection criterion for some adjustable parameters in the KLFs is developed to reduce the unreliability caused by experience. Furthermore, since only a small part of the binary measurements need to be augmented, the computational burden of the proposed KLFs are much lower than that of the traditional centralized algorithms.
\end{itemize}
\textbf{Notations:} ${\mathbb{R}}^r$ and ${\mathbb{R}}^{r\times s}$ denote the $r$-dimensional and $r\times s$ dimensional Euclidean spaces, respectively. $\mathrm{E}\{\cdot\}$ denotes mathematical expectation, while $\mathrm{diag}\{\cdot\}$ stands for block diagonal matrix. $O$ is zero matrix and $I$ stands for identity matrix. $\mathrm{Tr}(\cdot)$ represents the trace of matrix. For a matrix $A\in{\mathbb{R}}^{r\times r}$, its eigenvalues are denoted by $\lambda_i(A)$, $i=1,2,\cdots,r$, where the largest eigenvalue is denoted as $\lambda_{max}(A)$. $d_i(A)$ stands for the $i$-th diagonal element of $A$ and $d_{max}(A)$ represents the largest diagonal element. For a matrix $B$, $B_i$ denotes $i$-th column of $B$. For $X\in{\mathbb{R}}^{r\times r}$, $Y\in{\mathbb{R}}^{r\times r}$, $X>Y$ and $X\geq Y$ respectively mean that $X-Y$ is a positive definite matrix and a semi-positive definite matrix.
\section{Problem Formulation}
Consider the following nonlinear dynamic system with binary measurements:
\begin{equation}
\left\{ \begin{array}{l}
x_k=f(x_{k-1},u_{k-1})+C_{k-1}w_{k-1}\\
y^i_k=
\left\{ \begin{array}{l}
1,\ z^i_k\geq\tau_i\\
0,\ z^i_k<\tau_i
\end{array} \right. \ i=1,2,\cdots m
\end{array} \right.
\label{2.B.1}
\end{equation}
where the sensed variable $z^i_k$ of binary sensor $i$ is given by
\begin{equation}
z^i_k=h^i(x_k)+E^i_kv^i_k.\nonumber
\end{equation}
Here, $x_k\in {\mathbb{R}}^n$ is the system state, $y^i_k\in \mathbb{R}^1$ is the $i$th binary measurement. $\tau^i$ is the threshold of the binary sensor $i$, which is a constant. $u_{k-1}$ is the control input. $f(\cdot)\in {\mathbb{R}}^n$ and $h(\cdot)\in {\mathbb{R}}^1$ are arbitrary nonlinear functions, $C_{k-1}$ and $E^i_k$ are matrices with appropriate dimensions. $w_k$, $v^i_k$ and $v^j_k$, $i\neq j$ are uncorrected Gaussian white noises, and their covariances are $Q_k$, $R^i_k$ and $R^j_k$ respectively. Moreover, when the system state $x_k$ and the sensed variable $z^i_k$ are both linear, the system (\ref{2.B.1}) reduces to
\begin{equation}
\left\{ \begin{array}{l}
x_k=A_{k-1}x_{k-1}+B_{k-1}u_{k-1}+C_{k-1}w_{k-1}\\
y^i_k=
\left\{ \begin{array}{l}
1,\ z^i_k\geq\tau_i\\
0,\ z^i_k<\tau_i
\end{array} \right. \ i=1,2,\cdots m
\end{array} \right.
\label{2.B.2}
\end{equation}
where
\begin{equation}
z^i_k=D^i_kx_k+E^i_kv^i_k.\nonumber
\end{equation}
Here, $A_{k-1}$, $B_{k-1}$ and $D^i_k$ are matrices with appropriate dimensions.

As can be seen from the definition of $y^i_k$ in (1) and (2), binary sensors hardly provide valid information from outputs of 1 bit. Therefore, a novel uncertain measurement model will be developed to extract useful information from binary sensors.
To this end, let us define
\begin{equation}
\bar{y}^i_k\triangleq
\left\{ \begin{array}{l}
1,\ \bar{z}^i_k\geq\tau_i\\
0,\ \bar{z}^i_k<\tau_i
\end{array} \right. \ i=1,2,\cdots m.
\end{equation}
where $\bar{z}^i_{k|k-1}$ is the one-step prediction of sensed variable $z^i_k$. By comparing the definitions of $y^i_k$ and $\bar{y}^i_k$, we know that, when  $y^i_k\neq\bar{y}^i_k$, threshold $\tau^i$ will inevitably fall between $\bar{z}^i_k$ and $z^i_k$, which can be modeled as
\begin{align}
\tau^i=(0.5-\delta^i_k)\bar{z}^i_k+(0.5+\delta^i_k)z^i_k\ \ i\in\mathcal{I}_k,
\label{2.B.3}
\end{align}
where $\delta^i_k\in(-0.5,0.5)$ is an uncertain parameter whose exact value is unknown. $\mathcal{I}_k$ represents the index of the binary sensors whose output $y^i_k\neq\bar{y}^i_k$, which can be denoted as
\begin{align}
\mathcal{I}_k=\{i|\bar{y}^i_k\neq y^i_k\}.
\label{3.A.3}
\end{align}
Notice that the model (4) represents the threshold as a convex combination of $\bar{z}^i_k$ and $z^i_k$, and thus it can effectively extract the intrinsic measurement information of binary sensors.

Consequently, the purpose of this paper is to design KLFs for systems (1) and (2) based on the model (4) such that the mean-square errors of KLFs are minimal at each time.

\textbf{Remark 1:}
It is well known that the KLFs correct the one-step prediction after receiving the measurements from sensors, and thus obtain state estimate. In fact, the correction is needed because measurements contain new information that differs from the one-step prediction. This new information is the part of the measurement that really plays a role in the filter and is often referred to as innovation. According to (1) and (2), we know that the information contained in the binary measurement $y^i_k=1$ is $z^i_k\geq\tau^i$. Then, if the one-step prediction $\bar{z}^i_k\geq\tau^i$ (i.e. $\bar{y}^i_k=1)$, the information contained in the binary measurement will overlap with that of the one-step prediction. In this case, only little innovation is contained in $y^i_k$. Moreover, the same conclusion can be obtained at $y^i_k=\bar{y}^i_k=0$. Therefore, when the binary measurements do not belong to $\mathcal{I}_k$, they have little effect on the filter. On the contrary, those binary measurements belonging to $\mathcal{I}_k$, which contain more innovations, play a major
role in the filter. Based on this idea, it is proposed to extract the useful information from the binary measurements that belong to $\mathcal{I}_k$, and then the innovation-based model (4) is developed in this paper.

\textbf{Remark 2:} When $y^i_k$ changes (i.e., $y^i_k\neq y^i_{k-1}$), one can deduce that the threshold $\tau^i$ must fall into the interval between $z^i_{k-1}$ and $z^i_k$. To describe this phenomenon, a model was proposed in [19-21] as follows:
\begin{equation}
\tau^i=(0.5-\epsilon^i_k)z^i_{k-1}+(0.5+\epsilon^i_k)z^i_k,\ i\in\mathcal{S}_k,
\label{2.B.3}
\end{equation}
where $\epsilon^i_k\in(-0.5,0.5)$ was an uncertain parameter and $\mathcal{S}_k=\{i|y^i_{k-1}\neq y^i_k\}$. Though the model (6) is reasonable, it only captures the switching information in the binary measurements instead of the innovation. In fact, it has been pointed out that the innovation of the measurement plays a major role in the KLFs. Under this case, model (4) is considered to capture more information that is useful for the KLF than model (6).
\section{Main Results}
Before giving the main results of this paper, the number of sensors belonging to $\mathcal{I}_k$ is first defined as $m_k$, and
\begin{numcases}
{}z_{\mathcal{I},k}\triangleq\left[\begin{matrix}
   z^{i_{k1}}_k\ \cdots\ z^{i_{km_k}}_k
   \end{matrix}\right]^T\
\bar{z}_{\mathcal{I},k}\triangleq\left[\begin{matrix}
   \bar{z}^{i_{k1}}_k\ \cdots\ \bar{z}^{i_{km_k}}_k
   \end{matrix}\right]^T\nonumber\\
\tau_{\mathcal{I},k}\triangleq\left[\begin{matrix}
   \tau^{i_{k1}}\ \cdots\ \tau^{i_{km_k}}
   \end{matrix}\right]^T\
v_{\mathcal{I},k}\triangleq\left[\begin{matrix}
   v^{i_{k1}}_k\ \cdots\ v^{i_{km_k}}_k
   \end{matrix}\right]^T\nonumber\\
D_{\mathcal{I},k}\triangleq\left[\begin{matrix}
   (D^{i_{k1}}_k)^T\ \cdots\ (D^{i_{km_k}}_k)^T
   \end{matrix}\right]^T\nonumber\\
E_{\mathcal{I},k}\triangleq\mathrm{diag}(E^{i_{k1}}_k,\cdots,E^{i_{km_k}}_k)\ \ \ i_{k1},\cdots,i_{km_k}\in\mathcal{I}_k\nonumber\\
h_{\mathcal{I},k}(\cdot)\triangleq\left[\begin{matrix}
   h^{i_{k1}}(\cdot)\ \cdots\ h^{i_{km_k}}(\cdot)
   \end{matrix}\right]^T\nonumber\\
R_{\mathcal{I},k}\triangleq\mathrm{diag}(R^{i_{k1}}_k,\cdots,R^{i_{km_k}}_k)\nonumber\\
\Delta_{\mathcal{I},k}\triangleq\mathrm{diag}(\delta^{i_{k1}}_k,\cdots,\delta^{i_{km_k}}_k)
\end{numcases}

\subsection{Linear Kalman-Like Filter under Binary Sensors}
In this section, the Kalman-like filtering problem for system (2) will be solved. To this end, $\mathcal{I}_k$ needs to be determined first. For system (2), the one-step prediction $\bar{z}^i_k$ of sensed variable $z^i_k$ can be calculated as
\begin{equation}\begin{aligned}
\bar{z}^i_k=D^i_k\bar{x}_k,\ i=1,2,\cdots,m
\end{aligned}\end{equation}
where the one-step state prediction $\bar{x}_k$ is given by
\begin{equation}\begin{aligned}
\bar{x}_k=A_{k-1}\hat{x}_{k-1}+B_{k-1}u_{k-1}.
\end{aligned}\end{equation}
Then, $\mathcal{I}_k$ can be easily determined by (3), (5) and (8). Augmenting these binary measurements that belong to $\mathcal{I}_k$, one has
\begin{equation}\begin{aligned}
z_{\mathcal{I},k}=D_{\mathcal{I},k}x_k+E_{\mathcal{I},k}v_{\mathcal{I},k},
\end{aligned}\end{equation}
\begin{equation}\begin{aligned}
\bar{z}_{\mathcal{I},k}=D_{\mathcal{I},k}\bar{x}_k,
\end{aligned}\end{equation}
\begin{equation}\begin{aligned}
\tau_{\mathcal{I},k}=(0.5I-\Delta_{\mathcal{I},k})\bar{z}_{\mathcal{I},k}+(0.5I+\Delta_{\mathcal{I},k})z_{\mathcal{I},k}.
\end{aligned}\end{equation}
where $z_{\mathcal{I},k}$, $\bar{z}_{\mathcal{I},k}$, $v_{\mathcal{I},k}$, $E_{\mathcal{I},k}$, $D_{\mathcal{I},k}$, $\Delta_{\mathcal{I},k}$ and $\tau_{\mathcal{I},k}$ are defined in (7).
As can be seen from (12) that the threshold $\tau_{\mathcal{I},k}$ is represented as a linear transformation of $z_{\mathcal{I},k}$, and hence the one-step prediction of (12) can be given by
\begin{equation}\begin{aligned}
\bar{\tau}_{\mathcal{I},k}=(0.5I-\Delta_{\mathcal{I},k})\bar{z}_{\mathcal{I},k}+(0.5I+\Delta_{\mathcal{I},k})\bar{z}_{\mathcal{I},k}=\bar{z}_{\mathcal{I},k}.
\end{aligned}\end{equation}
Treating (11) as the measurement equation, the state estimate $\hat{x}_k$ for system (2) can be constructed as the following KLF structure:
\begin{equation}
\left\{ \begin{array}{l}
\hat{x}_k=\bar{x}_k+G^L_{\mathcal{I},k}(\tau_{\mathcal{I},k}-\bar{z}_{\mathcal{I},k})\\
\bar{x}_k=A_{k-1}\hat{x}_{k-1}+B_{k-1}u_{k-1}\\
\bar{z}_{\mathcal{I},k}=D_{\mathcal{I},k}\bar{x}_k.
\end{array} \right.
\label{2.B.2}
\end{equation}
where $G^L_{\mathcal{I},k}$ is the filter gain to be designed.

Substituting (12) into (14), the estimation error $\tilde{x}_k=x_k-\hat{x}_k$ of KLF (14) can be expressed as
\begin{equation}\begin{aligned}
\tilde{x}_k=&x_k-\bar{x}_k-
G^L_{\mathcal{I},k}(0.5I+\Delta_{\mathcal{I},k})(z_{\mathcal{I},k}-\bar{z}_{\mathcal{I},k}).
\label{3.B.7}
\end{aligned}\end{equation}
Obviously, due to the uncertainty $\Delta_{\mathcal{I},k}$ contained in $\tilde{x}_k$, the exact value of the estimation error covariance $\hat{P}_k=\mathrm{E}[(x_k-\hat{x}_k)(x_k-\hat{x}^T_k)]$ cannot be obtained. Therefore, a conservative estimation error covariance $\hat{\Phi}_k$ (i.e., an upper bound of $\hat{P}_k$) that incorporates all possible values of the uncertainty will be derived, and then the filter gain $G^L_{\mathcal{I},k}$ can be given by minimizing $\mathrm{Tr}(\hat{\Phi}_k)$ in Theorem 1.

\textbf{Theorem 1:} When $\mathcal{I}_k\neq\varnothing$, the upper bound $\hat{\Phi}_k$ of $\hat{P}_k$ that satisfies $\hat{\Phi}_k\geq\hat{P}_k$ for all $\Delta_k$ is calculated by
\begin{equation}\begin{aligned}
\hat{\Phi}_k=&0.25G^L_{\mathcal{I},k}[D_{\mathcal{I},k}\Upsilon_kD^T_{\mathcal{I},k}+\beta_kI+\Xi_k](G^L_{\mathcal{I},k})^T\\
&-0.5\Upsilon_kD^T_{\mathcal{I},k}(G^L_{\mathcal{I},k})^T-0.5G^L_{\mathcal{I},k}D_{\mathcal{I},k}\Upsilon_k+\Upsilon_k
\label{eq:3.C.7}
\end{aligned}\end{equation}
where $\alpha_k$ and $\beta_k$ are the given adjustable parameters satisfying
\begin{equation}\begin{aligned}
\alpha_k I>\Psi_k,\ \beta_k I>D_{\mathcal{I},k}\bar{\Phi}_kD^T_{\mathcal{I},k},
\label{eq:3.C.10}
\end{aligned}\end{equation}
and
\begin{numcases}
{}\Upsilon_k\triangleq\bar{\Phi}_k+\bar{\Phi}_kD^T_{\mathcal{I},k}(\beta_k I-D_{\mathcal{I},k}\bar{\Phi}_kD^T_{\mathcal{I},k})^{-1}D_{\mathcal{I},k}\bar{\Phi}_k\nonumber\\
\Xi_k\triangleq\Psi_k+\Psi_k(\alpha_k I-\Psi_k)^{-1}\Psi_k+\alpha_k I\\
\bar{\Phi}_k\triangleq A_{k-1}\hat{\Phi}_{k-1}A^T_{k-1}+C_{k-1}Q_{k-1}C^T_{k-1}\nonumber\\
\Psi_k\triangleq E_{\mathcal{I},k}R_{\mathcal{I},k}E^T_{\mathcal{I},k}\nonumber
\end{numcases}

Meanwhile, by minimizing $\mathrm{Tr}(\hat{\Phi}_k)$, the filter gain $G^L_{\mathcal{I},k}$ is obtained by
\begin{equation}\begin{aligned}
G^L_{\mathcal{I},k}=2\Upsilon^T_kD^T_{\mathcal{I},k}[D_{\mathcal{I},k}\Upsilon_kD^T_{\mathcal{I},k}+\beta_kI+\Xi_k]^{-1}.
\label{eq:3.C.9}
\end{aligned}\end{equation}
Furthermore, by minimizing the upper bound of the $\hat{\Phi}_k$ given by (16) and (19), the optimal $\alpha_k$ can be chosen as
\begin{equation}\begin{aligned}
\alpha_k=2d_{max}(\Psi_k).
\label{eq:3.C.11}
\end{aligned}\end{equation}

\textbf{Proof:} Substituting (2), (10) and (14) into (15), one has
\begin{align}
\tilde{x}_k=&A_{k-1}\tilde{x}_{k-1}+C_{k-1}w_{k-1}-G^L_{\mathcal{I},k}(0.5I+\Delta_{\mathcal{I},k})\nonumber\\
&\times(D_{\mathcal{I},k}(A_{k-1}\tilde{x}_{k-1}+C_{k-1}w_{k-1})+E_{\mathcal{I},k}v_{\mathcal{I},k})\nonumber\\
=&[I-0.5G^L_{\mathcal{I},k}D_{\mathcal{I},k}-0.5G^L_{\mathcal{I},k}2\Delta_{\mathcal{I},k}D_{\mathcal{I},k}]A_{k-1}\tilde{x}_{k-1}\nonumber\\
&+[I-0.5G^L_{\mathcal{I},k}D_{\mathcal{I},k}-0.5G^L_{\mathcal{I},k}2\Delta_{\mathcal{I},k}D_{\mathcal{I},k}]C_{k-1}w_{k-1}\nonumber\\
&-0.5G^L_{\mathcal{I},k}(I+2\Delta_{\mathcal{I},k})E_{\mathcal{I},k}v_{\mathcal{I},k}.
\end{align}
Then, the estimation error covariance is calculated by
\begin{equation}\begin{aligned}
\hat{P}_k=&(I-0.5G^L_{\mathcal{I},k}D_{\mathcal{I},k}-0.5G^L_{\mathcal{I},k}2\Delta_{\mathcal{I},k}D_{\mathcal{I},k})\\
&\times(A_{k-1}\hat{P}_{k-1}A^T_{k-1}+C_{k-1}Q_{k-1}C^T_{k-1})\\
&\times(I-0.5G^L_{\mathcal{I},k}D_{\mathcal{I},k}-0.5G^L_{\mathcal{I},k}2\Delta_{\mathcal{I},k}D_{\mathcal{I},k})^T\\
&+0.25G^L_{\mathcal{I},k}(I+2\Delta_{\mathcal{I},k})\Psi_k(I+2\Delta_{\mathcal{I},k})^T(G^T_{\mathcal{I},k})^L
\label{eq:3.C.12}
\end{aligned}\end{equation}
where $\Psi_k$ is defined in (18).

Notice that $2\Delta_{\mathcal{I},k}2\Delta_{\mathcal{I},k}\leq I$ and $\hat{\Phi}_{k-1}$ is an upper bound for $\hat{P}_{k-1}$. In this case, it follows from Lemma 1 in [20] that the following inequalities hold:
\begin{equation}\begin{aligned}
&(I-0.5G^L_{\mathcal{I},k}D_{\mathcal{I},k}-0.5G^L_{\mathcal{I},k}2\Delta_{\mathcal{I},k}D_{\mathcal{I},k})\\
&\times(A_{k-1}\hat{P}_{k-1}A^T_{k-1}+C_{k-1}Q_{k-1}C^T_{k-1})\\
&\times(I-0.5G^L_{\mathcal{I},k}D_{\mathcal{I},k}-0.5G^L_{\mathcal{I},k}2\Delta_{\mathcal{I},k}D_{\mathcal{I},k})^T\\
\leq& N_k\bar{\Phi}_kD^T_{\mathcal{I},k}(\beta_k I-D_{\mathcal{I},k}\bar{\Phi}_kD^T_{\mathcal{I},k})^{-1} D_k\bar{\Phi}_kN_k^T\\
&+N_k\bar{\Phi}_kN_k^T+0.25\beta_k G^L_{\mathcal{I},k}(G^L_{\mathcal{I},k})^T,
\label{eq:3.C.14}
\end{aligned}\end{equation}
\begin{equation}\begin{aligned}
&0.25G^L_{\mathcal{I},k}(I+2\Delta_{\mathcal{I},k})\Psi_k(I+2\Delta_{\mathcal{I},k})^T(G^L_{\mathcal{I},k})^T\\
\leq&0.25G^L_{\mathcal{I},k}[\Psi_k+\Psi_k(\alpha_k I-\Psi_k)^{-1}\Psi_k+\alpha_k I](G^L_{\mathcal{I},k})^T,
\label{eq:3.C.15}
\end{aligned}\end{equation}
where $N_k=I-0.5G^L_{\mathcal{I},k}D_{\mathcal{I},k}$, $\bar{\Phi}_k$ is defined in (18), $\alpha_k$ and $\beta_k$ are the given parameters that satisfy the conditions in (17).

Substituting (23) and (22) into (22), one can deduce that $\hat{\Phi}_k\geq\hat{P}_k$ holds for all $\Delta_k$. In this case, the optimization objective is chosen as $\mathrm{Tr}(\hat{\Phi}_k)$, and taking the partial derivative of $\mathrm{Tr}(\hat{\Phi}_k)$ with respect to $G^L_{\mathcal{I},k}$ yields that
\begin{equation}\begin{aligned}
\frac{\partial \mathrm{Tr}(\hat{\Phi}_k)}{\partial G^L_{\mathcal{I},k}}=-\Upsilon^T_kD^T_{\mathcal{I},k}+0.5G^L_{\mathcal{I},k}[D_{\mathcal{I},k}\Upsilon_kD^T_{\mathcal{I},k}+\beta_kI+\Xi_k]
\label{eq:3.C.16}\nonumber
\end{aligned}\end{equation}
where $\Upsilon_k$ and $\Xi_k$ are defined in (18). Let $\partial \mathrm{Tr}(\hat{\Phi}_k)/\partial G^L_{\mathcal{I},k}$ equal to $O$, the filter gain $G^L_{\mathcal{I},k}$ can be obtained from (19).

Next, an optimal way of selecting $\alpha_k$ will be given by minimizing the upper bound of $\hat{\Phi}_k$. Substituting (19) into (16), $\hat{\Phi}_k$ is rearranged into
\begin{equation}\begin{aligned}
\hat{\Phi}_k=&\Upsilon_k-\Upsilon_kD^T_{\mathcal{I},k}(D_{\mathcal{I},k}\Upsilon_kD^T_{\mathcal{I},k}+\beta_kI+\Xi_k)^{-1}D_{\mathcal{I},k}\Upsilon_k.
\nonumber
\end{aligned}\end{equation}
Notice that $\Xi_k$ is a diagonal matrix, and thus one has
\begin{equation}\begin{aligned}
\hat{\Phi}_k\leq&-\Upsilon_kD^T_{\mathcal{I},k}(D_{\mathcal{I},k}\Upsilon_kD^T_{\mathcal{I},k}+\beta_kI+d_{max}(\Xi_k)I)^{-1}D_{\mathcal{I},k}\\
&\times\Upsilon_k+\Upsilon_k.
\nonumber
\end{aligned}\end{equation}
Obviously, only the matrix $\Xi_k$ contains the parameter $\alpha_k$. Then, to minimize the upper bound of $\hat{\Phi}_k$, the objective function should be
\begin{equation}
\left\{ \begin{array}{l}
\min\limits_{\alpha_k} d_{max}(\Xi_k)\\
\mathrm{s.t.}\ \alpha_kI>\Psi_k
\end{array} \right..
\label{eq:3.B.23}
\end{equation}

To solve (25), the maximum diagonal element of $\Xi_k$ needs to be determined. Then, it follows from the definition of $\Xi_k$ in (18) that
\begin{equation}\begin{aligned}
d_i(\Xi_k)=d_i(\Psi_k)+\frac{d^2_i(\Psi_k)}{\alpha_k-d_i(\Psi_k)}+\alpha_k,\ i=1,2,\cdots,m_k.
\nonumber
\end{aligned}\end{equation}
Taking the partial derivative of $d_i(\Xi_k)$ with respect to $d_i(\Psi_k)$, one has
\begin{equation}\begin{aligned}
\frac{\partial d_i(\Xi_k)}{\partial d_i(\Psi_k)}=1+\frac{d_i(\Psi_k)(2\alpha_k-d_i(\Psi_k))}{(\alpha_k-d_i(\Psi_k))^2}.
\nonumber
\end{aligned}\end{equation}
Then, notice that the constraint in (25) is equivalent to $\alpha_k>d_{max}(\Psi_k)$
in which case the above equation is greater than 0. Thus, we can know that $d_i(\Xi_k)$ increases as $d_i(\Psi_k)$ increases when $\alpha_k$ is invariable.
In this case, one can deduce that
\begin{equation}\begin{aligned}
d_{max}(\Xi_k)=d_{max}(\Psi_k)+\frac{d^2_{max}(\Psi_k)}{\alpha_k-d_{max}(\Psi_k)}+\alpha_k.
\nonumber
\end{aligned}\end{equation}

Finally, by taking the derivative of $d_{max}(\Xi_k)$ with respect to $\alpha_k$ and making it equal to $0$, the analytical solution of (25) can be obtained from (20). The proof is completed. $\square$

When $\mathcal{I}_k=\varnothing$, no innovations are included in the binary measurements, and thus the state estimate is equal to the one-step prediction:
\begin{equation}\begin{aligned}
\hat{x}_k=\bar{x}_k=A_{k-1}\hat{x}_{k-1}+B_{k-1}u_{k-1},\ \mathcal{I}_k=\varnothing.
\end{aligned}\end{equation}
In this case, one has
\begin{equation}\begin{aligned}
\hat{P}_k&=E[(x_k-\bar{x}_k)(x_k-\bar{x}_k)^T]\\
&\leq A_{k-1}\hat{\Phi}_{k-1}A^T_{k-1}+C_{k-1}Q_{k-1}C^T_{k-1}\\
&=\bar{\Phi}_k,\ \mathcal{I}_k=\varnothing.
\end{aligned}\end{equation}

Through the analysis in this section, the computation procedures for linear binary Kalman-like filter (LBKLF) can be summarized by Algorithm 1.
\begin{algorithm}
\caption{Linear Binary Kalman-Like Filter}
\begin{algorithmic}[1]\label{algo:1}
\STATE  Initialize: $k=0$, $\hat{x}_0$, $\hat{\Phi}_0$, $\tau^i$, $i=1,2,\cdots,m$;
\STATE  $k\gets k+1$;
\STATE  Input: $\hat{x}_{k-1}$, $\hat{\Phi}_{k-1}$ and $y^i_k$, $i=1,2,\cdots,m$;
\STATE  Calculate $\bar{x}_k$ and $\bar{\Phi}_k$ by (9) and (18);
\STATE  Calculate $\bar{z}^i_k$ and $\bar{y}^i_k$, $i=1,2,\cdots,m$ by (8) and (3);
\STATE  Determine $\mathcal{I}_k$ by (5);
\IF{$\mathcal{I}_k\neq\varnothing$}
\STATE  Calculate $\alpha_k$ by (20);
\STATE Determine $\beta_k$ by experience. A range for reference is $\lambda_{max}(D_{\mathcal{I},k}\bar{\Phi}_kD^T_{\mathcal{I},k})<\beta_k\leq 2\lambda_{max}(D_{\mathcal{I},k}\bar{\Phi}_kD^T_{\mathcal{I},k})$;
\STATE  Calculate filter gain $G^L_{\mathcal{I},k}$ by (19);
\STATE  Calculate the state estimate $\hat{x}_k$ and conservative estimation error covariance $\hat{\Phi}_k$ by (14) and (16), respectively;
\ELSE
\STATE  $\hat{x}_k=\bar{x}_k$, $\hat{\Phi}_k=\bar{\Phi}_k$;
\ENDIF
\STATE Return to step 2.
\end{algorithmic}
\end{algorithm}
\subsection{Nonlinear Kalman-Like Filter under Binary Sensors}
In this section we discuss the design method of KLF for the nonlinear dynamic system (1). Since both the state $x_k$ and sensed variable $z^i_k$ of system (1) are nonlinear, the one-step predictions cannot be obtained directly by linear transformations as in the previous section. A common way to deal with the above nonlinearity is linearizing $f(\cdot)$ and $h^i(\cdot)$ by using Taylor first-order expansion which is however
only applicable to differentiable and low nonlinear systems. In contrast, the unscented transform (UT) [3] calculates the statistical properties of the random variables through a specific set of sampling points, which is applicable to arbitrary nonlinear systems and performs well with moderate nonlinearity. Based on this fact, the UT is adopted in this paper.

For system (1), the one-step state prediction $\bar{x}_k$, error covariance $\bar{P}_k=E[(x_k-\bar{x}_k)(x_k-\bar{x}_k)^T]$ and one-step prediction $\bar{z}^i_k$ of sensed variable $z^i_k$ can be calculated by using UT:
\begin{equation}
\bar{x}_k=\sum^{2n}_{j=0}w^m_jf(\hat{\chi}_{k-1,j},u_{k-1})
\end{equation}
\begin{align}
\bar{P}_k=&\sum^{2n}_{j=0}w^c_j(f(\hat{\chi}_{k-1,j},u_{k-1})-\bar{x}_k)(f(\hat{\chi}_{k-1,j},u_{k-1})-\bar{x}_k)^T\nonumber\\
&+C_{k-1}Q_kC^T_{k-1},
\label{eq:3.B.4}
\end{align}
\begin{equation}
\bar{z}^i_k=\sum^{2n}_{j=0}w^m_jh^i(\bar{\chi}_{k,j})\ i=1,2,\cdots,m.
\end{equation}
where $w^m_j$ and $w^c_j$ are the weights in the UT, $\hat{\chi}_{k-1,j}$ and $\bar{\chi}_{k,j}$ are the sigma points in the UT and their specific expressions are presented in Appendix. Then, according to (3), (5) and (30), $\mathcal{I}_k$ can be determined.

When $\mathcal{I}_k=\varnothing$, the state estimate $\hat{x}_k$ and conservative estimation error covariance $\hat{\Phi}_k$ are equal to the one-step predictions:
\begin{equation}\begin{aligned}
\hat{x}_k=\bar{x}_k,\ \hat{\Phi}_k=\bar{P}_k,\ \mathcal{I}_k=\varnothing.\nonumber
\label{eq:3.B.2}
\end{aligned}\end{equation}

When $\mathcal{I}_k\neq\varnothing$, similar to the previous section, augmenting the binary measurements that belong to $\mathcal{I}_k$, one has
\begin{equation}\begin{aligned}
z_{\mathcal{I},k}=h_{\mathcal{I},k}(x_k)+E_{\mathcal{I},k}v_{\mathcal{I},k},
\nonumber
\end{aligned}\end{equation}
\begin{equation}\begin{aligned}
\bar{z}_{\mathcal{I},k}=\sum^{2n}_{j=0}w^m_jh_{\mathcal{I},k}(\bar{\chi}_{k,j}),
\nonumber
\end{aligned}\end{equation}
\begin{equation}\begin{aligned}
\tau_{\mathcal{I},k}=(0.5I-\Delta_{\mathcal{I},k})\bar{z}_{\mathcal{I},k}+(0.5I+\Delta_{\mathcal{I},k})z_{\mathcal{I},k},
\end{aligned}\end{equation}
where $h_{\mathcal{I},k}(\cdot)$ is defined in (7). Then, the one-step prediction of (31) can also be given by
\begin{equation}\begin{aligned}
\bar{\tau}_{\mathcal{I},k}=(0.5I-\Delta_{\mathcal{I},k})\bar{z}_{\mathcal{I},k}+(0.5I+\Delta_{\mathcal{I},k})\bar{z}_{\mathcal{I},k}=\bar{z}_{\mathcal{I},k}.
\end{aligned}\end{equation}
Thus, the nonlinear KLF can be constructed for system (1):
\begin{equation}
\left\{ \begin{array}{l}
\hat{x}_k=\bar{x}_k+G^N_{\mathcal{I},k}(\tau_{\mathcal{I},k}-\bar{z}_{\mathcal{I},k})\\
\bar{x}_k=\sum^{2n}_{j=0}w^m_jf(\hat{\chi}_{k-1,j},u_{k-1})\\
\bar{z}_{\mathcal{I},k}=\sum^{2n}_{j=0}w^m_jh_{\mathcal{I},k}(\bar{\chi}_{k,j})
\end{array} \right.
\label{2.B.2}
\end{equation}
where $G^N_{\mathcal{I},k}$ is the filter gain that will be designed in the Theorem 2.

\textbf{Theorem 2:}  When $\mathcal{I}_k\neq\varnothing$, the conservative estimation error covariance $\hat{\Phi}_k$ of the KLF (33) that incorporates all possible values of the uncertainty $\Delta_{\mathcal{I},k}$ is calculated by
\begin{equation}\begin{aligned}
\hat{\Phi}_k=&\bar{P}_k-0.5\bar{P}^{xz}_k(G^N_{\mathcal{I},k})^T-0.5G^N_{\mathcal{I},k}(\bar{P}^{xz}_k)^T\\
&+0.25G^N_{\mathcal{I},k}[\bar{P}^{zz}_k+\bar{P}^{zz}_k(\varepsilon_kI-\bar{P}^{zz}_k)^{-1}\bar{P}^{zz}_k\\
&+(\varepsilon_k+\xi_k)I](G^N_{\mathcal{I},k})^T+\frac{1}{\xi_k}\bar{P}^{xz}_k(\bar{P}^{xz}_k)^T.
\label{eq:3.B.9}
\end{aligned}\end{equation}
where $\varepsilon_k$ and $\xi_k$ are given adjustable parameters satisfying
\begin{equation}\begin{aligned}
\varepsilon_k I>\bar{P}^{zz}_k,\ \xi_k>0,
\label{eq:3.B.11}
\end{aligned}\end{equation}
and
\begin{equation}\begin{aligned}
\bar{P}^{xz}_k=\sum^{2n}_{j=0}w^c_j(\bar{\chi}_{k,j}-\bar{x}_k)(h_{\mathcal{I},k}(\bar{\chi}_{k,j})-\bar{z}_{\mathcal{I},k})^T,
\label{eq:3.B.13}
\end{aligned}\end{equation}
\begin{equation}\begin{aligned}
\bar{P}^{zz}_k=&\sum^{2n}_{j=0}w^c_j(h_{\mathcal{I},k}(\bar{\chi}_{k,j})-\bar{z}_{\mathcal{I},k})(h_{\mathcal{I},k}(\bar{\chi}_{k,j})-\bar{z}_{\mathcal{I},k})^T\\
&+E_kR_kE^T_k.
\label{eq:3.B.14}
\end{aligned}\end{equation}
Meanwhile, by minimizing $\mathrm{Tr}(\hat{\Phi}_k)$, the nonlinear filter gain $G^N_{\mathcal{I},k}$ can be obtained by
\begin{equation}\begin{aligned}
G^N_{\mathcal{I},k}=2\bar{P}^{xz}_k[\bar{P}^{zz}_k+\bar{P}^{zz}_k(\varepsilon_k I-\bar{P}^{zz}_k)^{-1}\bar{P}^{zz}_k+(\varepsilon_k+\xi_k) I]^{-1}
\label{eq:3.B.10}
\end{aligned}\end{equation}
Furthermore, when minimizing the upper bound of the $\hat{\Phi}_k$ given by (34) and (38), the optimal $\varepsilon_k$ is chosen as
\begin{equation}\begin{aligned}
\varepsilon_k=2\lambda_{max}(\bar{P}^{zz}_k).
\label{eq:3.B.12}
\end{aligned}\end{equation}

\textbf{Proof:} Substituting (31) into (33), the estimate error $\tilde{x}_k=x_k-\hat{x}_k$ is given by
\begin{equation}\begin{aligned}
\tilde{x}_k=x_k-\bar{x}_k-G^N_{\mathcal{I},k}(0.5+\Delta_{\mathcal{I},k})(z_{\mathcal{I},k}-\bar{z}_{\mathcal{I},k}).
\nonumber
\end{aligned}\end{equation}
Then, the estimation error covariance is calculated by
\begin{equation}\begin{aligned}
\hat{P}_k=&E[\tilde{x}_k\tilde{x}^T_k]\\
=&\bar{P}_k-0.5\bar{P}^{xz}_k(I+2\Delta_{\mathcal{I},k})(G^N_{\mathcal{I},k})^T\\
&-0.5G^N_{\mathcal{I},k}(I+2\Delta_{\mathcal{I},k})(\bar{P}^{xz}_k)^T\\
&+0.25G^N_{\mathcal{I},k}(I+2\Delta_{\mathcal{I},k})\bar{P}^{zz}_k(I+2\Delta_{\mathcal{I},k})(G^N_{\mathcal{I},k})^T
\label{eq:3.B.15}
\end{aligned}\end{equation}
where the expressions for $\bar{P}^{xz}_k=E[(x_k-\bar{x}_k)(z_k-\bar{z}_k)^T]$ and $\bar{P}^{zz}_k=E[(z_k-\bar{z}_k)(z_k-\bar{z}_k)^T]$ can be given by using UT as shown in (36) and (37).

Using Lemma 1 in [20], one has
\begin{equation}\begin{aligned}
&G^N_{\mathcal{I},k}(I+2\Delta_{\mathcal{I},k})\bar{P}^{zz}_k(I+2\Delta_{\mathcal{I},k})(G^N_{\mathcal{I},k})^T\\
\leq& G^N_{\mathcal{I},k}[\bar{P}^{zz}_k+\bar{P}^{zz}_k(\varepsilon_k I-\bar{P}^{zz}_k)^{-1}\bar{P}^{zz}_k+\varepsilon_k I](G^N_{\mathcal{I},k})^T.
\label{eq:3.B.16}
\end{aligned}\end{equation}
Moreover, it follows from Lemma 2.2 in [22] that
\begin{equation}\begin{aligned}
&-\bar{P}^{xz}_k\Delta_{\mathcal{I},k}(G^N_{\mathcal{I},k})^T-(G^N_{\mathcal{I},k})^T\Delta_{\mathcal{I},k}(\bar{P}^{xz}_k)^T\\
\leq& 0.25\xi_kG^N_{\mathcal{I},k}(G^N_{\mathcal{I},k})^T+\frac{1}{\xi_k}\bar{P}^{xz}_k(\bar{P}^{xz}_k)^T,
\label{eq:3.B.17}
\end{aligned}\end{equation}
where $\varepsilon_k$ and $\xi_k$ are given adjustable parameters that satisfy the conditions in (35). Then, substituting (41) and (42) into (40), we know that $\hat{\Phi}_k\geq\hat{P}_k$ holds for all $\Delta_k$,  where the expression
of $\hat{\Phi}_k$ is shown in (34). In this case, the $\hat{\Phi}_k$ can be seen as a conservative estimation error covariance of the KLF (33), and the effect of the approximation error caused by UT can also be included in $\hat{\Phi}_k$.

Then, calculating $\partial\mathrm{Tr}(\hat{\Phi}_k)/\partial G^N_{\mathcal{I},k}$ and make it equal to $O$, the nonlinear filter gain $G^N_{\mathcal{I},k}$ is given by (38).

Substituting (38) into (34), $\hat{\Phi}_k$ is rearranged into
\begin{equation}\begin{aligned}
\hat{\Phi}_k=&\bar{P}_k-\bar{P}^{xz}_k[\bar{P}^{zz}_k+M(\varepsilon_k)+\xi_kI]^{-1}(\bar{P}^{xz}_k)^T\\
&+\frac{1}{\xi_k}\bar{P}^{xz}_k(\bar{P}^{xz}_k)^T,
\label{eq:3.B.19}
\end{aligned}\end{equation}
where $M(\varepsilon_k)=\bar{P}^{zz}_k(\varepsilon_k I-\bar{P}^{zz}_k)^{-1}\bar{P}^{zz}_k+\varepsilon_kI$ represents the terms associated with $\varepsilon_k$. Obviously, for (43), the following inequality holds:
\begin{align}
\hat{\Phi}_k\leq&\bar{P}_k-\bar{P}^{xz}_k[\bar{P}^{zz}_k+\lambda_{max}(M(\varepsilon_k))I+\xi_kI]^{-1}(\bar{P}^{xz}_k)^T\nonumber\\
&+\frac{1}{\xi_k}\bar{P}^{xz}_k\bar{P}^{xz}_k.
\nonumber
\end{align}
Thus, to minimize the upper bound of $\hat{\Phi}_k$, the objective function should be
\begin{equation}
\left\{ \begin{array}{l}
\min\limits_{\varepsilon_k} \lambda_{max}(M(\varepsilon_k))\\
\mathrm{s.t.}\ \varepsilon_kI>\bar{P}^{zz}_k
\end{array} \right..
\label{eq:3.B.23}
\end{equation}
Using the basic properties of matrix eigenvalues, it is not difficult to prove that the eigenvalues of $M(\varepsilon_k)$ are
\begin{equation}\begin{aligned}
\lambda_i(M(\varepsilon_k))=\frac{\lambda_i^2(\bar{P}^{zz}_k)}{\varepsilon_k-\lambda_i(\bar{P}^{zz}_k)}+\varepsilon_k,\ i=1,2,\cdots,m_k.
\nonumber
\end{aligned}\end{equation}

Finally, using the similar approach as for solving (24) in Theorem 1, the analytic solution of (43) can be obtained from (38). The proof is completed. $\square$

According to the analysis in this section, the computation procedures for nonlinear binary Kalman-like filter (NBKLF) are summarized by Algorithm 2.
\begin{algorithm}
\caption{Nonlinear Binary Kalman-Like Filter}
\begin{algorithmic}[1]\label{algo:2}
\STATE  Initialize: $k=0$, $\hat{x}_0$, $\hat{\Phi}_0$, $\tau^i$, $i=1,2,\cdots,L$;
\STATE  $k\gets k+1$;
\STATE  Input: $\hat{x}_{k-1}$, $\Phi_{k-1}$ and $y^i_k$, $i=1,2,\cdots,L$;
\STATE  Calculate $\bar{x}_k$ and $\bar{P}_k$ by (28) and (29), respectively;
\STATE  Calculate $\bar{z}^i_k$ and $\bar{y}^i_k$ $i=1,2,\cdots,L$ by (30) and (3);
\STATE  Determine $\mathcal{I}_k$ by (5);
\IF{$\mathcal{I}_k\neq\varnothing$}
\STATE  Calculate $\varepsilon_k$ by (39);
\STATE  Determine $\xi_k$ which often takes values in the range $0<\xi_k\leq2\mathrm{Tr}(\bar{P}^{xz}_k(\bar{P}^{xz}_k)^T)$;
\STATE  Calculate nonlinear filter Gain $G^N_{\mathcal{I},k}$ by (38);
\STATE  Calculate the state estimate $\hat{x}_k$ and conservative estimation error covariance $\hat{\Phi}_k$ by (33) and (34), respectively;
\ELSE
\STATE  $\hat{x}_k=\bar{x}_k$, $\hat{\Phi}_k=\bar{\Phi}_k$;
\ENDIF
\STATE Return to step 2;
\end{algorithmic}
\end{algorithm}


\textbf{Remark 3:}
Notice that, Lemma 1 in [20] and Lemma 2.2 in [22] are two matrix inequalities that commonly used to deal with the uncertainty, but the adoption of these two matrix inequalities necessarily introduces adjustable parameters which are often chosen empirically in most literatures. In fact, experience is sometimes unreliable. Therefore, to reduce the influence of experience on the filter as much as possible, an optimal selection of the adjustable parameters $\alpha_k$ and $\varepsilon_k$ are given by minimizing the upper bound of $\hat{\Phi}_k$ in this paper. However, the DFKLF in [20] does not take this into account, although similar adjustable parameters also exist in it.

\textbf{Remark 4:} When constructing KLF based on (6), one-step prediction of (6) needs to be calculated:
\begin{equation}
\bar{\tau}^i=(0.5-\epsilon^i_k)\hat{z}^i_{k-1}+(0.5+\epsilon^i_k)\bar{z}^i_k,\ i\in\mathcal{S}_k,
\label{2.B.3}
\end{equation}
where $\hat{z}_{k-1}$ is the estimation of $z_{k-1}$, and it is obtained from
$\hat{x}_{k-1}$ and the measurement equations. Unfortunately, due to uncertainty $\epsilon^i_k$, the exact value of $\bar{\tau}^i$ is unknown. Thus, $\epsilon^i_k$ has to be ignored, and then the following KLF can be constructed for single binary sensor [20-21] :
\begin{equation}
\hat{x}_k=\bar{x}_k+G^i_k(\tau^i-0.5\hat{z}^i_{k-1}-0.5\bar{z}^i_{k}).\\
\end{equation}
The formula (46) is the structure of KLFs in [20-21]. Obviously, their estimation performance are reduced because $\epsilon^i_k$ is ignored. In addition, the uncertainties caused by binary sensors were also ignored in [19] when constructing the moving horizon estimator, and the specific analysis of this can be found in Remark 1 of [21]. In contrast, benefiting from the form of innovation-based model (4), the uncertainty $\Delta_{\mathcal{I},k}$ is offset rather than ignored when calculating $\bar{\tau}_{\mathcal{I},k}$ in (13) and (32). In this case, the LBKLF and NBKLF have better estimation performance than the methods in [20-21].

\textbf{Remark 5:} The algorithms proposed in this paper take centralized approach, i.e. augmenting the measurements. However, it follows from (7) that only those binary measurements containing innovations are augmented. In this case,
the computational complexity of LBKLF and NBKLF are $O(n^3+n^2m_k+nm_k^2+m_k^3)$, where $m_k$ represents the number of binary measurements contained in the set $\mathcal{I}_k$. Obviously, the computational complexity of the proposed algorithms is not directly related to the total number of binary sensors $m$, but it is related to $m_k$ which is less than $m$. Therefore, the proposed algorithms overcome the disadvantage that the computational complexity of traditional centralized approach increases sharply when the number of sensors increasing.
\section{Simulation Results}
\subsection{$O_2$ Content Estimation in Arteries}
Consider arterial $O_2$ content estimation using the noninvasive binary pulmonary sensors where the physiological model for the arterial $O_2$ content is [20]:
\begin{equation}\begin{aligned}
&x_{k+1}=fx_k+U_k+w_k\\
&U_k=(1-f)(1.34Hb+0.003(au_k+c_ke_k))-f\mu\\
&a=P_{ATM}-P_{H_2O}\\
&c_k=[1-u_k(1-RQ)]/RQ
\nonumber
\end{aligned}\end{equation}
where $x_k$ is the arterial $O_2$ content, $u_k$ is the percentage of $O_2$ in the inhaled air and is set by surgeons, and thus it can be considered as the control input. $f$ represents the fraction of shunted blood. $e_k$ is the partial pressure of exhaled $CO_2$ and can be measured directly. $Hb$ is the amount of hemoglobin, $P_{ATM}$ and $P_{H_2O}$ are the atmospheric and water vapor pressures, $\mu$ reflects the patient-specific metabolic rate, $RQ$ is the respiratory quotient. Specifically, the constant parameters above are chosen as [24]: $f=0.75$, $Hb=12\ g/dL$, $P_{ATM}=760\ mmHg$, $P_{H_2O}=47\ mmHg$, $\mu=5\ mL/dL$, $RQ=0.8$. $u_k$ is set to $60\%$. Meanwhile, the sensed variable $z^i_k$ is constructed by three other inputs: tidal volume, respiratory rate and peak inspiratory, and it is proportional to the $O_2$ content [24]:
\begin{equation}\begin{aligned}
z^i_k=D^i_kx_k+v^i_k,\ i=1,2,\cdots,L.
\nonumber
\end{aligned}\end{equation}
where $D^i_k=0.5$, $i=1,2,\cdots,L$. $w_k$ and $v^i_k$ are Gaussian white noise with covariance $1$ and $0.02$, respectively. Here, the $O_2$ content change process is monitored by 10 binary sensors whose thresholds are set to $\tau^i=61+0.5i,\ i=1,2,\cdots,10$.

By implementing Algorithm 1, the trajectories of true arterial $O_2$ content and the estimated arterial $O_2$ content by using the LBKLF are plotted in Fig. 1, which shows that the proposed LBKLF can estimate the arterial $O_2$ content well. Due to the random noises, the estimation performance is assessed by the root mean square error (RMSE), and 100 Monte Carlo runs are implemented to approximate the ideal RMSE. Then, the RMSEs of the LBKLF, DFKLF in [20] and MHE in [19] are plotted in Fig. 2, where the sliding window size for the MHE is chosen to be 100 and the fusion criterion for the DFKLF is chosen to be fast covariance intersection fusion [23]. It can be seen from Fig. 2 that the estimation accuracy of the LBKLF is higher than that of DFKLF and MHE, which is mainly caused by three factors: i). The uncertainties caused by the binary sensors are offset in LBKLF, rather than ignored directly as in DFKLF and MHE. ii). Compared with the model (6) in [19-21], the proposed innovation-based model (4) can capture more innovations which play a major role in the filer;  iii). Compared with the DFKLF, the proposed LBKLF gives an optimal selection criterion for the adjustable parameter $\alpha_k$, and thus reducing the unreliability caused by experience.

On the other hand, Fig. 3 shows that under 100 Monte Carlo simulations, the number of binary sensors to be augmented (i.e., $m_k$) is much smaller than the total number of binary sensors $m=10$, and combining this fact with the Remark 4 means that the computational complexity of LBKLF is low. Meanwhile, to show this point more intuitively, the computational overheads of different methods are listed in Tab. I, from which we can see that the LBKLF is more computationally efficient than DFKLF and MHE.
\begin{figure}[thpb]
      \centering
      \includegraphics[scale=0.6]{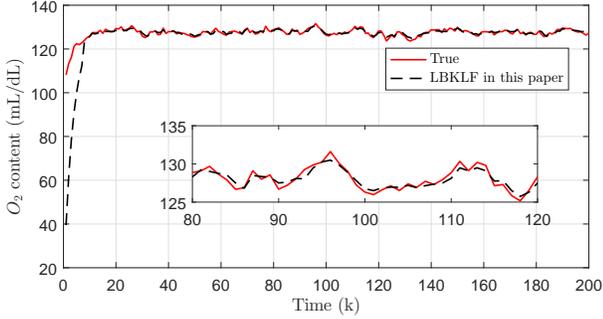}
      \caption{The true $O_2$ content and the estimated $O_2$ content by using LBKLF.}
      \label{O2_content}
    \end{figure}
   \begin{figure}[thpb]
      \centering
      \includegraphics[scale=0.6]{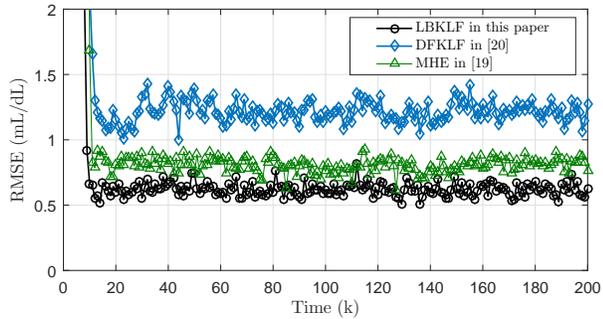}
      \caption{Comparisons of the $O_2$ content RMSEs of the LBKLF in this paper, the DFKLF in [20] and the MHE in [19].}
      \label{O2_RMSE}
    \end{figure}
\begin{figure}[thpb]
      \centering
      \includegraphics[scale=0.6]{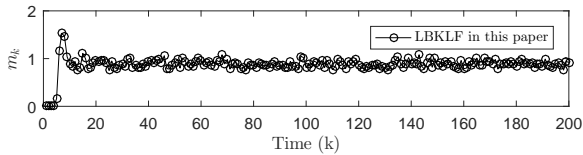}
      \caption{The average number of binary sensors belonging to $\mathcal{I}_k$ in 100 Monte Carlo simulations.}
\end{figure}

\subsection{Nonlinear Numerical Simulation}
Consider the following nonlinear state-space equation:
\begin{equation}
x_{k+1}=\left[\begin{matrix}
   x_{k+1,1}\\
   x_{k+1,2}
   \end{matrix}\right]
   =\left[\begin{matrix}
   g(x_{k,1})+0.1g(x_{k,2})\\
   g(x_{k,2})+0.1g(x_{k,1})
   \end{matrix}\right]+U_k+w_k\nonumber
\end{equation}
where
\begin{equation}\begin{aligned}
&g(x)\triangleq0.9x+\frac{x+100}{x^2+1},\
U_k=\left[\begin{matrix}
   2\mathrm{cos}(k/5)\\
   2\mathrm{sin}(k/5)
   \end{matrix}\right].\nonumber
\end{aligned}\end{equation}
Then, 18 binary sensors are used to observe $x_{k}$, and their sensed variables $z^i_k$ ($i=1,2,\cdots,18$) measure the logarithmic distance between state $x_{k,i}$ $(i=1,2)$ and specific values, which can be expressed as
\begin{equation}
z^i_k=\left\{ \begin{array}{l}
\ln(\sqrt{(x_{k,1}-15-2\times i)^2})+v^i_k\ \ \ \ i=1,2\cdots,9,\\
\ln(\sqrt{(x_{k,2}+22-3.5\times i)^2})+v^i_k\ \ i=10,\cdots,18,
\end{array} \right.\nonumber
\end{equation}
and their thresholds are set to
\begin{equation}
\tau^i_k=\left\{ \begin{array}{l}
\ln(0.5)=-0.69\ \ \ \ \ \ \ i=1,2\cdots,9,\\
\ln(0.875)=-0.13\ \ \ \ i=9,10\cdots,18.
\end{array} \right.\nonumber
\end{equation}
$w_k$ and $v^i_k$ are Gaussian white noise with covariance $\mathrm{diag}(0.09,0.25)$ and $0.01$, respectively.

By implementing Algorithm 2, the true trajectories and the estimated trajectories by using NBKLF are plotted in Fig. 4, from which we can see that the NBKLF tracks the true trajectories well. Meanwhile, 100 Monte Carlo runs are performed to approximate the theoretical RMSE, which is shown in Fig. 5. It is seen from this figure that the RMSE of NBKLF is maintained at a low level when estimating the nonlinear system. Notice that the DFKLF and MHE in [19-20] are only applicable to linear dynamic systems and therefore NBKLF is not compared with them. On the other hand, though 18 binary sensors were used to observe the system state, it is shown from Fig. 6 that only 1 binary sensor contained useful innovation at each moment on average. This allows the NBKLF to be run with a low cost of $5\times10^{-4}s$ per moment on average.


\begin{table}
\caption{Average computation overhead per moment of the LBKLF in this paper,the DFKLF in [20] and the MHE in [19]}
\begin{center}
\begin{tabu} to 0.47\textwidth{X[5,c]|X[c]|X[c]|X[c]}
\hline
Algorithms  &LBKLF &DFKLF &MHE\\
\hline
Computational overhead ($10^{-5}$s)    &$1.6$       &$13.1$           &$19.6$\\
\hline
\end{tabu}
\end{center}
\end{table}
\begin{figure}[thpb]
      \centering
      \includegraphics[scale=0.6]{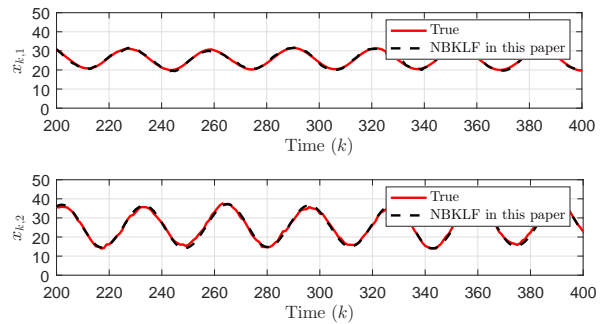}
      \caption{The true trajectories and the estimated trajectories by using NBKLF.}
      \label{nonlinear_trajectories}
    \end{figure}
   \begin{figure}[thpb]
      \centering
      \includegraphics[scale=0.6]{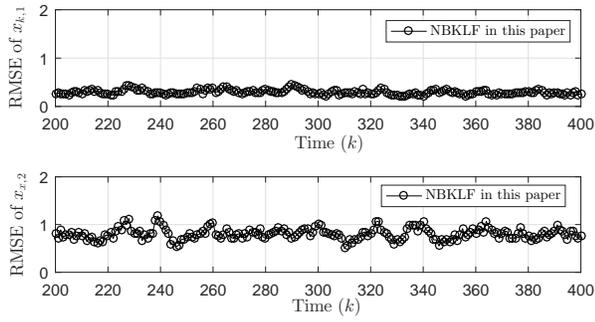}
      \caption{The RMSEs of the NBKLF in this paper.}
      \label{nonlinear_RMSE}
    \end{figure}
   \begin{figure}[thpb]
      \centering
      \includegraphics[scale=0.6]{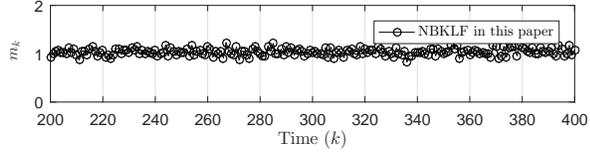}
      \caption{The average number of binary sensors belonging to $\mathcal{I}_k$ in 100 Monte Carlo simulations.}
      \label{nonlinear_sensor_content}
    \end{figure}

\section{Conclusion}
In this paper, a novel uncertainty measurement model for binary sensors was developed such that the innovations within the binary outputs could be captured.
When considering linear dynamic systems, a conservative estimation error covariance with adjustable parameters was first derived by matrix inequality, and then the filter gain was given by minimizing the trace of this estimation error covariance. Particularly, the optimal selection of an adjustable parameter was developed by minimizing the upper bound of the estimation error covariance. Following the similar idea, a KLF was also designed for nonlinear dynamic systems by using unscented transform. In addition, the computational effort of the proposed algorithms were kept low since only a small number of binary measurements need to be augmented.
Finally, two illustrative examples are employed to show the effectiveness and advantages of the proposed methods.
\appendix
For system (1),
the sampling strategy of $\hat{\chi}_{k-1,j}$ is
\begin{equation}\begin{aligned}
\hat{\chi}_{k-1,j}=\hat{\sigma}_{k-1,j}+\hat{x}_{k-1},\ j=0,1,\cdots,2n\nonumber
\end{aligned}\end{equation}
\begin{equation}\begin{aligned}
\hat{\sigma}_{k-1,j}=
\left\{ \begin{array}{l}
O\ \ \ \ \ \ \ \ \ \ \ \ \ \ \ \ \ \ \ \ \ \ \ \ j=0\\
-(\sqrt{(n+\eta)\hat{\Phi}_{k-1}})_j\ \ j=1,2\cdots,n\\
(\sqrt{(n+\eta)\hat{\Phi}_{k-1}})_{j-n}\ j=n+1,n+2,\dots,2n
\end{array} \right.\nonumber
\end{aligned}\end{equation}
where $\sqrt{A}$ denotes the Cholesky decomposition of a positive definition matrix $A$. Then, the predictions $\bar{x}_k$ and $\bar{P}_k$ can be computed as in (28) and (29). The sigma point $\bar{\chi}_{k,j}$ can be given by
\begin{equation}\begin{aligned}
\bar{\chi}_{k,j}=\bar{\sigma}_{k,j}+\bar{x}_k,\ j=0,1,\cdots,2n\nonumber
\end{aligned}\end{equation}
\begin{equation}\begin{aligned}
\bar{\sigma}_{k,j}=
\left\{ \begin{array}{l}
O\ \ \ \ \ \ \ \ \ \ \ \ \ \ \ \ \ \ \ \ \ \ \ j=0\\
-(\sqrt{(n+\eta)\bar{P}_k})_j\ \ \ \ j=1,2\cdots,n\\
(\sqrt{(n+\eta)\bar{P}_k})_{j-n}\ \ \ j=n+1,n+2,\dots,2n
\end{array} \right.\nonumber
\end{aligned}\end{equation}
Moreover, the weights $w^m_j$ and $w^c_j$ are given by
\begin{equation}\begin{aligned}
\left\{ \begin{array}{l}
w^m_{j}=w^c_{j}=\frac{1}{2(n+\eta)},\ j=1,2,\cdots,2n, \nonumber\\
w^c_0=w^m_0+(1-a^2+b),\ w^m_0=\frac{1}{n+\eta}
\end{array} \right.\nonumber
\end{aligned}\end{equation}
where the constants are chosen as $\eta=a^2(n+\kappa)-n$, $b=2$, $\kappa=0$, $a=1$. The principle of the UT can be referred to [3], which will not be repeated here.

%





\ifCLASSOPTIONcaptionsoff
  \newpage
\fi



\end{document}